\documentclass[a4paper]{jpconf}

\usepackage{graphicx}
\usepackage{subcaption} 
\usepackage{algorithm}
\usepackage{algpseudocode}
\usepackage{amsmath}
\usepackage{amssymb}
\usepackage[numbers,sort&compress,square]{natbib}
\usepackage{svg}

\begin{document}

\title{Load constrained wind farm flow control through multi-objective multi-agent reinforcement learning}

\author{Teodor Åstrand$^{1}$, Marcus Binder Nilsen$^{1}$, Iasonas Tsaklis$^{1}$, Tuhfe Göçmen$^{1}$, Pierre-Elouan  Réthoré$^{1}$ and Nikolay Dimitrov$^{1}$}

\address{$^1$Department of Wind and Energy Systems, Technical University of Denmark, Roskilde, Denmark}

\ead{tobas@dtu.dk}

\begin{abstract}
This study presents a multi-agent reinforcement learning (MARL) framework for load-constrained wind farm flow control (WFFC). While wake steering can enhance total wind farm power, it significantly alters the fatigue loading environment of downstream turbines. To address this, we integrate an Independent Soft Actor-Critic (I-SAC) architecture with a data-driven, local inflow sector-averaged surrogate model to provide real-time estimates of Damage Equivalent Loads (DELs). By incorporating these estimates into a shaped reward function, turbine-specific agents are trained to maximise power production while adhering to specific load-increase thresholds ($\Delta_{max}$) of 10\%, 20\%, and 30\% relative to a baseline controller. The framework is implemented within the WindGym environment using the DYNAMIKS flow solver with Dynamic Wake Meandering (DWM) to model and capture non-stationary wake physics. Results indicate that the MARL agents successfully learn collaborative policies that prioritise power gain while actively retreating from high-DEL control strategies. 
\end{abstract}

\vspace{1em}

\section{Introduction}
A wind farm is a complex environment where multiple objectives, such as minimising wake losses, maximising power output, and mitigating loads, must be balanced to increase the energy yield and extend the lifetime of the turbines. However, current operating controllers are often limited to local single-objective greedy strategies. Although a common industry practice, these controllers neglect turbine interactions, which are crucial for a holistic approach to wind farm flow control (WFFC). As shown in \cite{THOMSEN1999121}, turbines operating in wakes of upstream turbines experience increased structural loads compared to corresponding free flow conditions. As turbine lifetime is a crucial factor in wind energy, there is an increasing focus on methods that consider loads in their control approach at the wind farm level.

Furthermore, WFFC is a field characterised by inherent uncertainty, as it involves finding optimal control strategies under turbulent wind conditions, measurement limitations, and aerodynamic complexity. Through the use of reinforcement learning (RL), robust policies can be learned under this given uncertainty without requiring an explicit model of wind farm aerodynamics. In this study, a multi-agent reinforcement learning (MARL) framework is integrated with a data-driven surrogate model to optimise total farm power while maintaining real-time constraints on damage equivalent loads (DELs). To study the direct impact of wake steering on structural loads, we focus on the blade root flapwise moment. Previous studies support this choice, as in \cite{Debusscher_2022}, the blade-root flapwise bending moment was shown to approach its constraint during wake-steering operation, while \cite{PADULLAPARTHI2022445} used blade-root fatigue as a key objective in multi-objective RL-based WFFC. Furthermore, the findings in \cite{damiani_assessment_2018} emphasise the asymmetric nature of blade loads; specifically, changing the sign of the yaw offset can either result in load alleviation or an increase. 

The sensitivity of blade-root flapwise moments to yaw misalignment highlights their relevance in a DEL-constrained optimal control. In addition, the potential for asymmetric load responses adds an additional layer of complexity, providing a challenging and relevant environment for testing the ability of RL to navigate multi-objective trade-offs.

RL is a strong contender for wake steering due to its ability to adapt to high-variance environments, learn nonlinear control behaviours, and exploit flow dynamics that may be challenging to model explicitly. However, the use of purely AI-driven controllers can raise concerns due to the difficulty of interpreting, explaining and predicting their behaviour in safety-critical settings. To address this, this study actively shapes the learned RL policies by coupling information from a DEL-based surrogate model into the reward formulation. From this, the objective is to learn policies that prioritise power production while actively retreating from high-DEL control strategies. 

\begin{figure}
    \centering
    \includegraphics[width=0.75\linewidth]{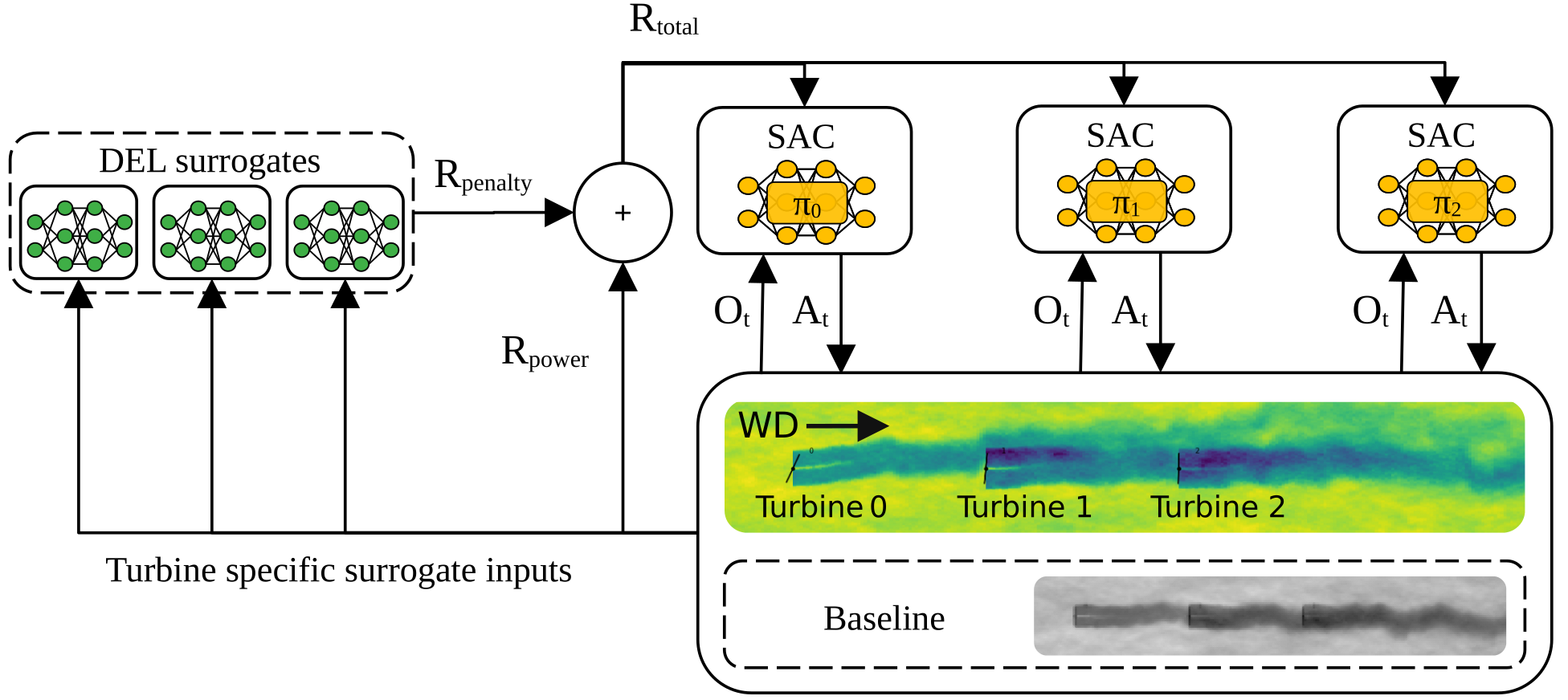}
    \caption{Overview of the utilised MARL setup using WindGym.} 
    \label{fig:setup_marl}
\end{figure}

The MARL strategy implemented in this work utilises a multi-agent architecture, as illustrated in Figure~\ref{fig:setup_marl}. In this framework, turbine-specific agents, governed by individual policies $\pi_i$, execute closed-loop control over their respective turbines. To optimise farm-wide performance, these agents must learn to coordinate their actions, as successful WFFC depends on the emergence of collective, collaborative policies. This coordination is facilitated through the use of a shared global reward signal, aligning individual actions with the farm's total power objectives and DEL constraints. 

\section{Background}
Single-agent RL has been shown to be a promising approach in the WFFC community \cite{GOCMEN2025115605}. Building on this, MARL has also proven to be an effective approach in several studies ~\cite[e.g.,][]{monroc2025wfcrlmultiagentreinforcementlearning, KADOCHE2023119129, PADULLAPARTHI2022445}, with one potential benefit being scalability through MARL approaches, which utilise centralised training and decentralised execution.

The choice of simulation environment is a critical factor for RL and WFFC. In \cite{KADOCHE2023119129}, a steady state environment was implemented. With this approach, one loses accuracy in how the wake is represented, since the wake propagates instantaneously to the downstream turbine(s). To improve upon the fidelity of the environment, a dynamic simulator can be utilised. This is employed in \cite{monroc2025wfcrlmultiagentreinforcementlearning}, which highlights that using a higher fidelity simulator increases the complexity of the problem significantly and comes with much higher computational costs. 


The current state of RL and WFFC mostly focuses on one objective, typically related to power production. However, although limited, there exist a few studies which include agents capable of learning policies considering multiple objectives, such as \cite{monroc2025wfcrlmultiagentreinforcementlearning} that penalises loads and \cite{PADULLAPARTHI2022445} that uses MARL to optimise power while mitigating loads. Based on these encouraging first results, we extend the analysis by including i) dynamic inflow, ii) a fully data-driven representation of the structural loads (via a DEL surrogate), iii) the capability to adjust the level of additional loading allowed under the constraint, and iv) leveraging a state-of-the-art multi-agent architecture to optimise collective wake steering WFFC.


\section{Problem formulation}
In this study, we strive to achieve optimal wake steering policies that maximise power $P$ for a wind farm with $N_{turbines}=3$ turbines by letting turbine specific agents control the yaw offset $\gamma_i$ for maximum wind farm power while constraining the increase in DEL of the blade root flap-wise moment for the turbine(s), with respect to baseline normal operation. 

The primary intention of this framework is to support the transition of RL-based WFFC toward field readiness by facilitating load-aware and load-constrained policies. By incorporating structural boundaries into the MARL training process, we address the requirement for structural reliability, which remains a primary barrier to adopting purely power-optimal AI controllers in the field.

The constraint is based on a 'max-to-max' comparison where we evaluate the maximum turbine-specific DEL during MARL control against the maximum turbine-specific DEL from the baseline greedy controller that controls solely for local power, unaware of the wake effect. Based on this framework, we aim to achieve optimal control policies targeting the following optimisation problem:


\begin{equation}
\begin{aligned}
\max_{\{\gamma_i\}} \quad & \sum_{i=1}^{N_{turbines}} P_i(\gamma_i) \\
\text{subject to} \quad & -30 \leq \gamma_i \leq 30, \quad \forall i \\
& 
\frac{\max(\text{DEL}_{\text{agent}})}{\max(\text{DEL}_{\text{baseline}})} \leq 1 + \Delta_{max}
\end{aligned}
\label{eq:objective}
\end{equation}
 Departing from the methodologies in 
 \cite{PADULLAPARTHI2022445} and \cite{monroc2025wfcrlmultiagentreinforcementlearning}, we introduce a novel approach to load integration in RL-based WFFC. Specifically, we investigate the impact of incorporating a DEL-based penalty at different levels of $\Delta_{max}$ as 10\%, 20\%, 30\%, and unconstrained. From this, we formulate the following research questions:

\begin{itemize}
    \item How can we leverage multi-objective MARL to develop a controller capable of optimising the constrained objective in Equation \ref{eq:objective}?
    \item What are the robust metrics to monitor and quantify the compliance of our AI-based approach to the implemented constraint(s)?
    \item How can we quantify the potential performance benefit of MARL-based wake steering WFFC in terms of power gain under structural load constraints?
\end{itemize}

\section{Methodology}
To facilitate an environment for learning optimal wake steering policies, we formulate the control problem as a partially observable Markov decision process (POMDP), leveraging high-performance computing (HPC) resources to enable scalable training.

\subsection{WFFC as a POMDP}
In this framework, an agent interacts with the environment in discrete time steps $t$ by emitting an action $a_t$ and receiving a reward $r_{t+1}$. While a standard Markov decision process (MDP) assumes the state $s_t$ captures all information relevant to the future \cite{sutton1998reinforcement}, the full aerodynamic state of a wind farm is not directly observable. Factors such as turbulent wind fields and wake positions make a complete state representation unobtainable. Consequently, we operate in a POMDP in which agents receive observations $o_t$ that are sampled from the true underlying state.

Given this uncertainty in the observations, it becomes difficult to predict a single optimal action. Through leveraging MARL, we learn a probability distribution in the form of turbine-specific policies $\pi(a \mid o)$. Here, actions $a$ are sampled based on observations $o$ (representing a partial view of the true state $s$) at each discrete time step $t$. This formulation accounts for uncertainty in the underlying wind conditions, as the goal is to learn a distribution of reasonable actions based on incomplete information.

The experimental setup utilises the WindGym environment \cite{WindGym2025}, an RL framework built on DYNAMIKS \cite{dynamiks}. From this environment, we can train and test our MARL-based approach by formulating the wake steering problem as a POMDP, where the agent receives observations $o_t$ from wind measurements, takes actions $a_t$ for yaw control, and receives rewards $r_{t+1}$ reflecting farm power and DEL-based penalties. The simulator provides transition data $(o_t, a_t, r_{t+1}, o_{t+1})$ that we use to train RL agents to learn policies $\pi(a \mid o)$ that map observations to yaw actions.

\begin{itemize}
\item \textbf{Reward function}
To train an agent capable of controlling the wind farm and optimising the objective in Equation \ref{eq:objective} the following reward function for power is used:

\begin{equation}
r_{\text{power}} = \frac{\overline{P}_{\text{agent}}}{\overline{P}_{\text{baseline}}} - 1
\end{equation}
where $\overline{P}_{\text{agent}}$ corresponds to the average farm power for the MARL-based controller, and $\overline{P}_{\text{baseline}}$ denotes the corresponding average for the baseline. To alter agent policies in such a way that they adapt to the DEL limit $\Delta_{\text{max}}$, the following reward shaping strategy is used: 

\begin{equation}
    r_{\text{penalty}} = 
    \begin{cases}
        (\Delta_{\text{max}} - \Delta) & \text{if } \Delta > \Delta_{\text{max}}\\[2mm]
        0 & \text{otherwise}
    \end{cases}
\end{equation}

where $\Delta$ corresponds to the percentage increase in DEL  over a 10-minute sliding window when comparing $\max(\text{DEL}_{\text{agent}})$ against $\max(\text{DEL}_{\text{baseline}})$. From this, $r_{\text{penalty}}$ serves as a penalty that is active only when the constraint is violated in terms of DEL increase. Throughout this study, lowercase $r$ denotes the instantaneous reward at each time step, whereas uppercase $R$ refers to the cumulative return that the agents aim to maximise. The total reward is calculated as a dimensionless scalar utility:
\begin{align}
        r_{\text{total}} &= \max\bigl( r_{\text{power}} + r_{\text{penalty}},\, -10 \bigr)
\end{align}

Here, the lower bound of -10 serves as a clipping mechanism. In deep reinforcement learning, excessively large negative rewards can lead to high-variance gradients; therefore, this bound is used to prevent destabilisation during training. This ensures that, while the agent is strongly discouraged from violating DEL constraints, the neural network's numerical updates remain stable.


\item \textbf{Observations:} To communicate with the environment, each turbine-specific agent samples observations from WindGym encompassing global flow conditions based on local turbine states. By limiting observations to what is obtainable from standard SCADA data with the intent to support field-deployable operation, the agent utilises dynamic, unsteady signals for reactive control alongside rolling averages for stationary estimation. In this configuration, agents receive estimates of global conditions ($ws_{\text{global}}$, $wd_{\text{global}}$) and their respective sliding means ($\overline{ws}_{\text{global}}$, $\overline{wd}_{\text{global}}$), calculated as averages across all turbines. Locally, the agent focuses on the current yaw offset ($\gamma$) and its sliding mean ($\overline{\gamma}$). From each step $t$, the following observation vector is obtained by each agent:

\begin{equation}
o_t = [ws_{\text{global}}, \overline{ws}_{\text{global}}, wd_{\text{global}}, \overline{wd}_{\text{global}}, \gamma, \overline{\gamma}]
\end{equation}

\item \textbf{Actions:} Each turbine-specific agent controls the turbine yaw misalignment $\gamma_i$ as its action. Control actions are applied at 10-second intervals, with a maximum yaw rate constrained to 0.25$^\circ$ per second. 

\end{itemize}

\subsection{Numerical environment and inflow parameters}
The training and evaluation scenarios are characterized by a fixed, fully aligned inflow with properties defined in Table~\ref{tab:numerical_setup}. The inflow follows a uniform wind profile across the rotor disk, allowing for the isolation of wake and steering effects from vertical shear. The flow physics are resolved using DYNAMIKS, which incorporates the Dynamic Wake Meandering (DWM) model to capture the non-stationary behaviour of turbine wakes. 

During the training, the three policies for the three turbines are deployed in a total of $N_{\text{env}} = 15$ parallel environments, each with its own turbulence box selected randomly from a pool of 60 boxes. Each realisation is generated using the Mann model. This setup improves data sampling efficiency and combats agent overfitting to a single turbulence realisation. For comparison, we run a corresponding baseline environment under the same conditions, allowing for a direct comparison of the RL agent's performance with that of the baseline controller. 

The emulated environment tracks several physical quantities for agent observations and reward calculations, including global and local flow conditions and turbine yaw offsets. Furthermore, sectoral inflow data is sampled to estimate the blade-root flapwise bending moment via a data-driven surrogate model, as detailed in Section \ref{sec:DEL}.

\begin{table}[h]
\caption{Numerical environment and inflow parameters.}\label{tab:numerical_setup}
\vspace{-10pt}
\begin{center}
\begin{tabular}{ll}
\br
Category & Description / Value \\
\mr
\textbf{Numerical Tools} & DYNAMIKS (DWM-based solver), Mann turbulence model \\
\textbf{Inflow Conditions} & $ws = 10$ m/s, $wd = 270^\circ$, $TI = 5\%$ (Uniform profile) \\
\textbf{Mann Turbulence Box} & Length scale $L = 33.6$ m, Flow length $L_x = 12,276$ m \\
\textbf{Farm Layout} & 3 turbines, inline configuration, $6D$ spacing \\
\textbf{Turbine Model} & Siemens SWT-2.3 (2.3 MW) \\
\textbf{Constraint Driving Load} & Blade-root flapwise bending moment (DEL) \\
\br
\end{tabular}
\end{center}
\end{table}

\subsection{Independent soft actor-critic in the context of WFFC}
With a WFFC problem formulated as a POMDP, each environment effectively resembles a Markov chain from which state transitions can be sampled in order to learn optimal policies for each turbine-specific agent. 

With \cite{huang2022cleanrl} as a starting point, the MARL approach used in this study leverages an implementation of Independent Soft Actor-Critic (I-SAC), a multi-agent adaptation of Soft Actor-Critic (SAC) \cite{haarnoja2018softactorcriticoffpolicymaximum}, which enables decentralised and collaborative RL-based control. A core concept in SAC is its maximum entropy framework, which explicitly encourages stochasticity in the learned policies. By promoting exploration and preventing premature convergence to suboptimal deterministic behaviours, this entropy-regularised formulation improves the overall stability of learning, which is deemed advantageous in a WFFC setting.

For each turbine-specific SAC agent, two key types of neural networks are utilised: 

\noindent\textbf{Actor Network} $\pi_{\phi_i}(a \mid o)$: Maps wind observations to a stochastic yaw control policy, outputting a probability distribution over yaw commands. The subscript $\phi_i$ denotes the trainable parameters of the $i$-th actor network.

\noindent\textbf{Q-Networks} $Q_{\theta_i}(s, a)$: Serves as a critic and evaluates the quality of an action $a$ for a given state by evaluating the expected cumulative reward for said action. Here, $\theta_i$ represents the trainable parameters of the $i$-th $Q$-network. To ensure training stability, target networks are maintained for each Q-network to provide slowly evolving reference values during the policy update process.
In this continuous control setting, the Q-function provides a quantitative measure of the quality of each yaw action, guiding the learning process towards actions that maximise the expected farm power production and ability to avoid DEL-based penalties based on the turbine yaw actions.


Furthermore, because the control task is continuous, environmental resets are implemented at fixed intervals. This allows the agents to navigate toward optimal set points following a forced reset to zero yaw offset, ultimately promoting training stability. The pseudo-code presented in Algorithm~\ref{alg:isac_windfarm} gives a core overview of the I-SAC implementation.

\begin{algorithm}
\footnotesize
\caption{I-SAC Algorithm for WFFC in WindGym}
\label{alg:isac_windfarm}
\begin{algorithmic}[1]

\State \textbf{Initialise :} actor networks, Q-networks and target networks for each agent $i$

\For{each iteration}
    \State \textbf{Environment Interaction:}
    \For{each environment step in $N_{env}=15$ parallel environments}
        \For{each agent $i$}
            \State $a_t^i \sim \pi_{\phi_i}(\cdot | o_t^i)$ \quad // Sample yaw action from policy
        \EndFor
        \State $s_{t+1}, r_{t+1} \sim \text{env}(a_t^1, \ldots, a_t^N)$ \quad // Execute actions in wind farm simulator
        \State $\mathcal{D} \leftarrow \mathcal{D} \cup \{(o_t,a_t, r_{t+1}, o_{t+1})\}$ \quad // Store transition
    \EndFor
    
    \State \textbf{Policy Updates:}
    \For{each gradient step}
        \For{each agent $i$}
            \State Sample batch $\mathcal{B} = \{o, a, r, o'\} \sim \mathcal{D}$
            
            \State \textbf{Q-Network Update}
            
            \State \textbf{Actor Update}
            
            \State \textbf{Target Network Update}
        \EndFor
    \EndFor
\EndFor

\end{algorithmic}
\end{algorithm}


\subsection{Using DEL surrogates to guide learning in dynamic wind farm operation}
\label{sec:DEL}


The utilised DEL surrogate is a three-layer network trained on data from design load case (DLC) aeroelastic simulations in HAWC2~\cite{HAWC2}, as described in \cite{wes-2026-45}, for Siemens SWT 2.3 turbine; where the input features are local flow quantities averaged over 10-minute intervals and across the four sectors of the rotor to correlate with the 10-minute structural load response of the turbine. 

The surrogate model incorporates spatial wind field variations by utilising wind speed and turbulence intensity sampled from four distinct rotor disk sectors (left, right, top, and bottom). This sector-based approach allows the neural network to account for the impact of partial wakes and non-uniform inflow when predicting the DELs. Additionally, the yaw setting of each turbine is provided as input. With this sectoral design, the surrogate functionality becomes favourable for this study. 

The surrogate expects time-averaged local flow quantities and directly provides 10-minute DEL values. To utilise it as feedback in our dynamic environment, we average the environmental input over a 10-minute sliding window. Additionally, we perform spatial averaging by calculating the mean of 15 simulated sensors in each respective sector that sample wind data from the dynamic simulator. From this spatial and temporal averaging, combined with the neural network's ability to evaluate quickly, we can obtain an estimate of the DEL of the blade root flapwise moment at each environmental step for each turbine in all parallel training simulations, which is then coupled to the agent training workflow. 


\section{Results}

\begin{figure} [H]
    \centering
    \includegraphics[width=0.9\linewidth]{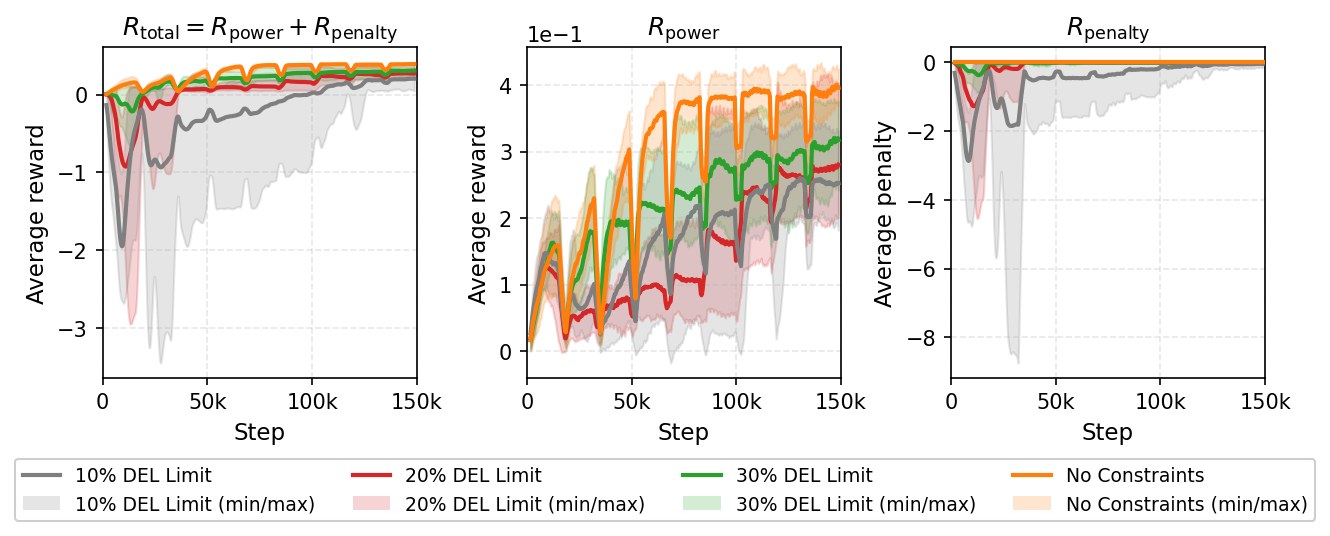}
    \caption{Agent training progress for six random seeds, each running in $N_{env}=15$ parallel environments for a cumulative 150,000 environment steps.}
    \label{fig:training}
\end{figure}

Agent versions were trained and evaluated for the different levels of $\Delta_{\max}$ (10\%, 20\%, 30\%, and unconstrained). Upon inspecting the training evolution in Figure \ref{fig:training}, the agents exhibit signs of learning, where both the constrained and unconstrained versions achieve higher power yield on average compared to the greedy baseline controller, due to the positive power reward. When observing the DEL-based penalty, it is apparent that agents generally learn to respect the constraints after approximately 30,000 steps, as the penalty's activity is zero for all versions apart from the 10\% DEL limit after this point. When comparing the power rewards for the unconstrained approach versus the penalised runs, it becomes clear that the unconstrained agents generally find higher power-yielding policies with a more stable learning curve when considering the power reward in Figure \ref{fig:training}. Upon reviewing the same figure, the environmental reset starting at step 20.000 becomes apparent, and when observing the recovery to optimal actions, recovery appears faster at each reset as training progresses for all variations. 

\begin{figure}[H]
    \centering
    \includegraphics[width=0.75\linewidth]{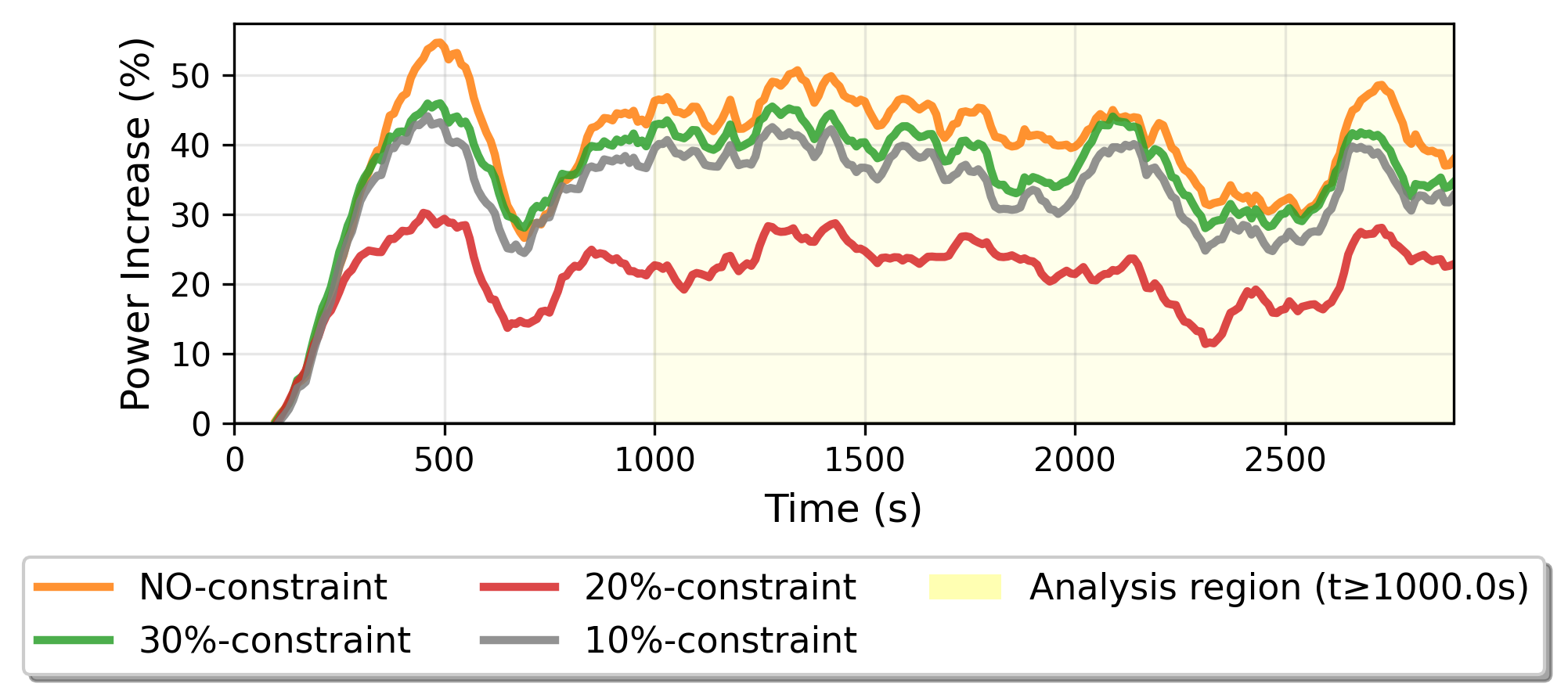}
    \caption{Time series of farm-level power production for the MARL-controlled case compared with the baseline greedy controller during a 3000-second simulation using a turbulence box unseen during training, for different levels of DEL constraint.}
    \label{fig:power}
\end{figure}
\begin{figure}[htbp]
    \centering
   
        \begin{subfigure}[b]{0.4\textwidth} 
        \centering
        \includegraphics[height=10.5cm]{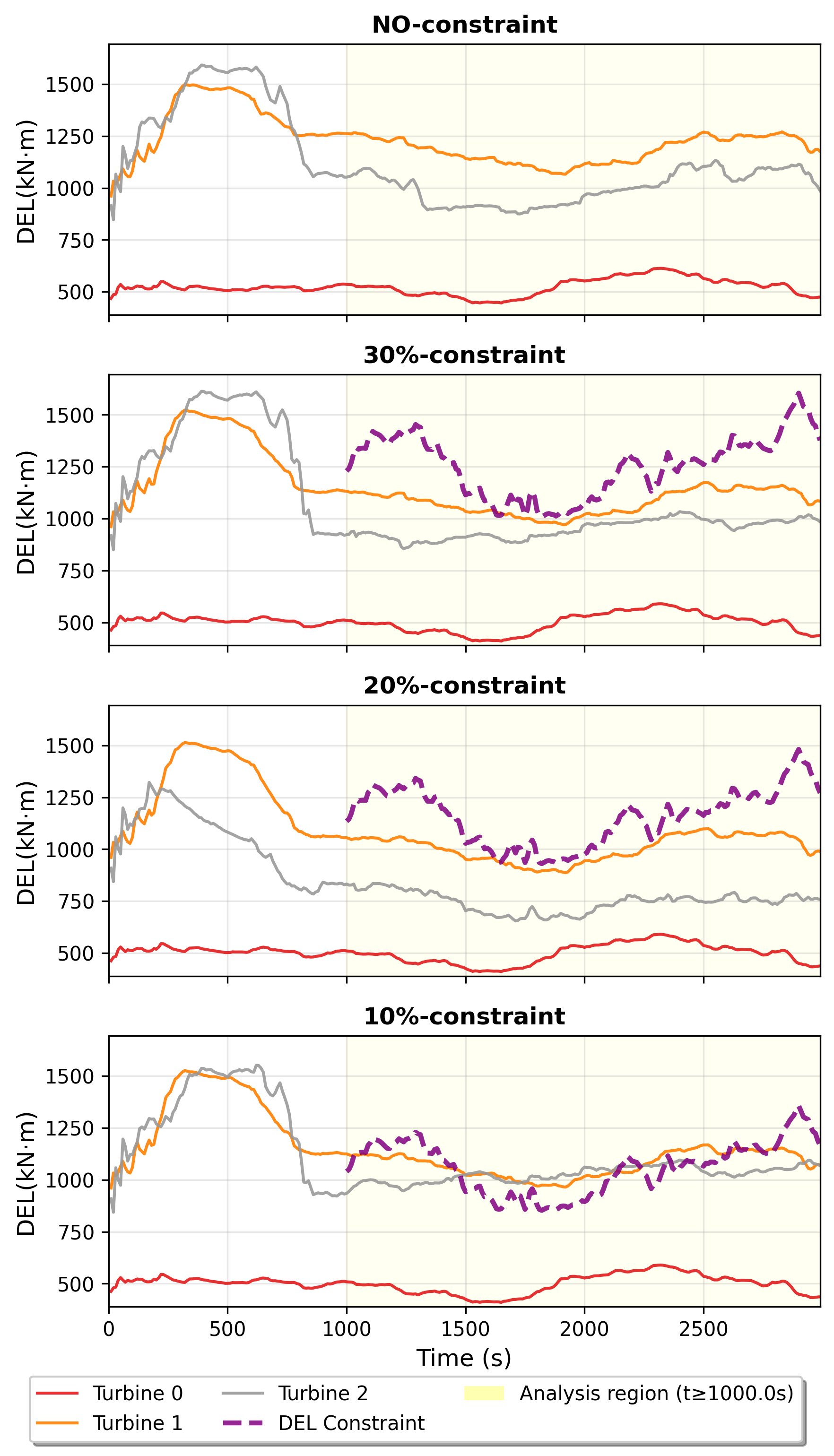}
        \caption{}
     \label{fig:load_time_series_sub}
    \end{subfigure}
        \begin{subfigure}[b]{0.4\textwidth} 
        \centering
        \includegraphics[height=10.5cm]{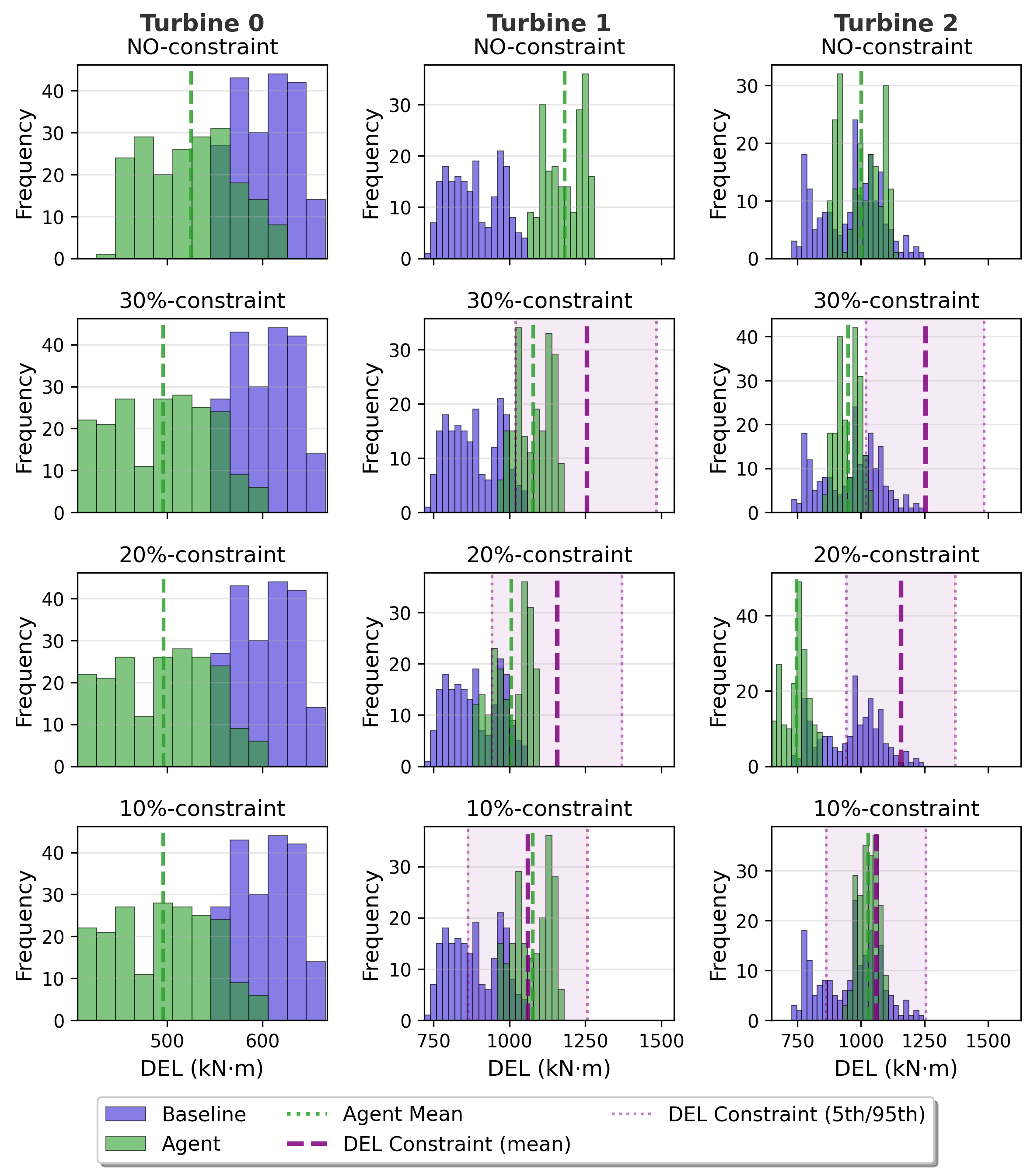}
        \caption{}
    \label{fig:load_hist_sub}
    \end{subfigure}
    \hspace{2.9cm}
    \caption{a) Time series of turbine-specific DEL for the blade root flap-wise moment, computed using a sliding-window method, for the MARL-controlled over a 3000-second simulation using a common unseen turbulence box. The DEL limit is based on the max DEL for all turbines from a baseline controller, where the analysis region represents the period after the flow is advected through the wind farm. b) Distribution of turbine-level DEL for the blade root flap-wise moment. The time-dependent DEL constraint is represented by its mean (dotted vertical line) and the $5^{th}$ and $95^{th}$ percentiles (shaded region).}
\end{figure}

For the evaluation phase, a single set of representative agents was selected for each constraint level to be deployed in a test environment featuring an unseen turbulence box. To assess their control performance and adaptability, all turbines were initialised at zero yaw offset, from which the agents were permitted to autonomously navigate the action space according to their learned policies $\pi$. As a benchmark, we compare our results to those of the baseline greedy controller. From this, we obtain the time series shown in Figures \ref{fig:power} and \ref{fig:load_time_series_sub}. Figure \ref{fig:power} shows that both the DEL-penalised and unconstrained RL agents achieve higher average farm power compared to the baseline greedy controller for all cases. Generally, the behaviour of the DEL-penalised agents follows the expected trend, where relaxed constraints correlate with higher power production; however, the agent subjected to the 10\% threshold serves as an outlier. In this instance, the policy outperforms the 20\% agent in terms of power but fails to strictly respect the constraint threshold, as evidenced in Figures \ref{fig:load_time_series_sub}. This outcome is consistent with the training dynamics shown in Figure \ref{fig:training}, where the DEL penalty for the 10\% threshold remains active significantly longer compared to the other thresholds. From this, one can suspect that these agents operate closer to the DEL threshold, and the policy becomes more susceptible to constraint violations when encountering an unseen turbulence realisation.



\begin{figure}
    \centering
    \includegraphics[width=0.9\linewidth]{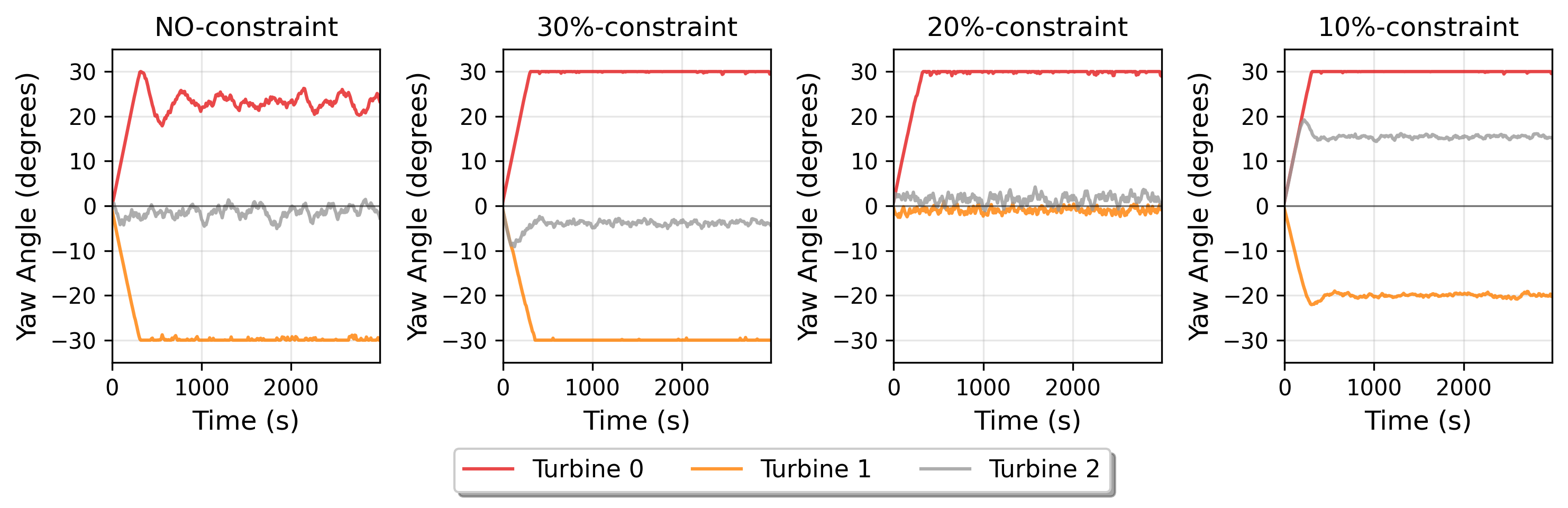}
    \caption{Yaw angles commanded by learned agent policies during evaluation.}
    \label{fig:yaw_angles}
\end{figure}

To investigate the shift in DEL, we use an analysis region as seen in Figure \ref{fig:load_time_series_sub} that begins 1000 seconds after the initial deployment. This allows the agents to apply their policies and gives the flow time to transition through the wind farm as the control actions generate changes in the wake. From this region, we generate the histograms in Figure \ref{fig:load_hist_sub}. 

Upon investigating DEL for the blade root flap-wise moment and its shift with the learned agent policies, we observe a general increase for the downstream turbine, Turbine 1, for the unconstrained agents. Furthermore, the asymmetric nature of blade loads is present in this scenario, as Turbine 0, the front-most turbine, exhibited a decreased blade root bending moment DEL compared to the baseline.  When observing the agent subjected to the DEL-based penalties, the same shift is not as apparent, as the policies manage to reduce the increase in DEL for the unseen turbulence box at 30\% and 20\% constraint thresholds for both Turbine 1 and 2, compared to the unconstrained agents. Specifically, the agents governed by these thresholds exhibit policies that actively attempt to remain below the dynamic load boundaries as seen in Figure \ref{fig:load_time_series_sub} and \ref{fig:load_hist_sub}. In contrast, for the version where the DEL penalty is enforced at the 10\% limit, the policy tends to fluctuate around the penalty line.

As illustrated by the commanded yaw angles in Figure \ref{fig:yaw_angles}, the 30\% constraint level demonstrates behaviour closely aligned with the unconstrained policy; however, the yaw offsets are shifted to prioritise lower DEL, which consequently results in reduced power production. For the agents constrained at 20\% and 10\%, the control strategy for Turbine 0 remains consistent with the unconstrained and 30\% cases. Significant deviations emerge in the yaw trajectories for Turbine 1, and at the 10\% constraint level, even Turbine 2 is actively engaged.


\section{Further discussion \& conclusions}
The adapted MARL framework successfully produces wake steering policies that balance increased power yield with structural load mitigation. By leveraging reward shaping with a common reward for all agents, collaborative policies are achieved that navigate the trade-off between energy production and the blade-root flapwise moment, effectively retreating from high-DEL control strategies at different thresholds. The evaluation confirms that the methodology enforces compliance with the imposed penalties, allowing for control over the shift in DEL relative to the baseline greedy controller. 

This is mainly supported by the metric from the evaluation in Figure \ref{fig:load_hist_sub}, which shows DEL distributions on turbines with and without WFFC, where the DEL penalised policies do not exhibit the same increase in DEL as the unconstrained policies. Additionally, the training data of the RL algorithm in Figure \ref{fig:training}, which depicts the DEL penalty, shows a reduction in penalty activity over time, indicating that the policies learn to respect the set threshold. 

Furthermore, quantifying the performance benefit for the unconstrained, 30\% and 20\% constrained cases reveals a clear trade-off during the learning phase (Figure \ref{fig:training}), where lower load thresholds typically lead to a more constrained power reward, as the agents are forced to prioritise penalty avoidance over maximising energy yield. However, in Figure \ref{fig:load_time_series_sub}, evaluation results demonstrate that although penalised policies successfully reduce DEL increase, the threshold can still be exceeded during deployment, as for the 10\% case. This behaviour can be attributed to the stochastic nature of the environment; while the agents learn to respect constraints across a range of training turbulence boxes, the frozen policy may not strictly satisfy the threshold when encountering an unseen turbulence box.

\subsection{Acknowledging uncertainties and proposal for future work}
While a time-averaged turbine response surrogate has inherent limitations in a dynamic environment, its primary purpose in this MARL framework is not high-fidelity accuracy, but rather to serve as a behavioural guide. The focus is to force agents to adapt their policies to the surrogate's dynamics, navigating the complex load trade-offs inherent in wake steering. By using temporal averaging, the objective shifts from precise DEL prediction to training policies that can interpret and respect structural boundaries under varying inflow. Although dynamic effects like yaw oscillations may be filtered out, the surrogate successfully shapes the agents' convergence toward safe yaw offsets. To address this limitation, a more traditional DEL computation method, such as rainflow counting, can be applied during evaluation.

Although the agent seemed to learn policies that strive to respect a limit to how much they are allowed to deviate from a DEL baseline, the final agents should ideally be evaluated for more scenarios with a mix of turbulence boxes and random seeds in order to support a better understanding of how safe the policies are in terms of maintaining a limited DEL increase. This is to ensure that the learned policies do not overfit to the pool of Mann turbulence boxes used, and to investigate the robustness of the policies.

Furthermore, a natural step toward achieving higher technological readiness for WFFC solutions involves transitioning to higher-fidelity simulations, such as Large-Eddy Simulations (LES) \cite{wes-7-2271-2022}. Future research should also incorporate multiple adversarial load channels, explore the edge case of 0\% permitted load increase, and extend the architecture to larger layouts to manage the increased complexity of multi-turbine wake interactions.

\ack
This study is supported by the IntelliWind project, funded through the EU MSCA Doctoral Networks programme with grant number 101168725. The support is greatly appreciated. The authors gratefully acknowledge the computational and data resources provided on the Sophia HPC Cluster at the Technical University of Denmark, DOI: 10.57940/FAFC-6M81.

{\footnotesize
\bibliographystyle{iopart-num}
\bibliography{Torque26_references}
}

\end{document}